\documentclass[PRL,aps,preprint,showpacs,amsmath,amssymb,superscriptaddress]{revtex4}
\usepackage[T1]{fontenc}
\usepackage[latin1]{inputenc}
\usepackage{amsmath}
\usepackage{graphicx}
\usepackage{epsfig}
\usepackage{bm}
\usepackage{dsfont}
\usepackage{color}

\newcommand{\nn}{\nonumber}

\newcommand{\bea}{\begin{eqnarray}}
\newcommand{\ea}{\end{eqnarray}}
\newcommand{\beq}{\begin{equation}}
\newcommand{\eq}{\end{equation}}
\newcommand{\bc}{\begin{center}}
\newcommand{\ec}{\end{center}}
\newcommand{\dg}{\dagger}
\newcommand{\la}{\langle}
\newcommand{\ra}{\rangle}
\newcommand{\ov}{\overline}
\newcommand{\partiel}[2]{\frac{\partial #1}{\partial #2}}

\begin{document}

\title{Temporal non-equilibrium dynamics of a Bose Josephson junction in presence of incoherent excitations}

\author{Mauricio Trujillo-Martinez}
\affiliation{Physikalisches Institut and Bethe Center for Theoretical Physics,
  Universit\"at Bonn, Nussallee, 12, D-53115 Bonn, 
Germany} 

\author{Anna Posazhennikova}
\email[Email: ]{anna.posazhennikova@rhul.ac.uk}

\affiliation{Department of Physics, Royal Holloway, University of London,
  Egham, Surrey TW20 0EX, United Kingdom}

\author{Johann Kroha}
\email[Email: ]{kroha@th.physik.uni-bonn.de}

\affiliation{Physikalisches Institut and Bethe Center for Theoretical Physics,
  Universit\"at Bonn, Nussallee, 12, D-53115 Bonn, 
Germany}

\date{\today}

\begin{abstract}
The time-dependent non-equilibrium dynamics of a Bose-Einstein condensate (BEC) typically generates incoherent excitations out of the condensate due to the finite frequencies present in the time evolution. We present a detailed derivation of a general non-equilibrium Green's function technique which describes the coupled time evolution of an interacting BEC and its single-particle excitations in a trap, based on an expansion in terms of the exact eigenstates of the trap potential. We analyze the dynamics of a Bose system in a small double-well potential with initially all particles in the condensate. When the trap frequency is larger than the Josephson frequency, $\Delta > \omega_J$, the dynamics changes at a characteristic time $\tau_c$ abruptly from slow Josephson oscillations of the BEC to fast Rabi oscillations driven by quasiparticle excitations in the trap. For times $t<\tau_c$ the Josephson oscillations are undamped, in agreement with experiments. We analyze the physical origin of the finite scale $\tau_c$ as well as its dependence on the trap parameter $\Delta$. 
\end{abstract}


\maketitle
\section{Introduction}

Time-dependent phenomena are at the heart of the physics of ultracold atomic
gases and, in particular, Bose-Einstein condensates (BEC). The creation of
BECs in optical traps \cite{Dalfovo99,Leggett01}  has opened a box of Pandora for the study of temporal
dynamics. This is because the relevant time scales in these systems allow for
direct access to the time evolution, like quench dynamics \cite{Greiner1,Greiner2}  or DC and AC
Josephson oscillations \cite{Albiez05,Levy07}, but also because most transport 
experiments are done by applying time-dependent external fields, like ratchet 
potentials \cite{Weitz09}, or by expansion experiments \cite{Dalfovo99,Leggett01}, since 
stationary transport cannot be driven easily in the neutral, finite-size systems. 

There are several powerful numerical methods which target dynamical behaviour of superfluids. For instance, in one-dimensional systems, quench and relaxation dynamics as well as temporal spread of correlations have been treated by 
time-dependent density matrix renormalization group \cite{Kollath07} and non-equilibrium dynamical mean-field theory for bosons \cite{Werner14}.
Also the exact quantum dynamics of a 1D bosonic Josephson junction (BJJ) has been investigated by solving the many-body Schr\"odinger equation of motion by using the multiconfigurational time-dependent Hartree for bosons (MCTDHB) method \cite{Sakmann09}. This work demonstrated that 
methods based on the standard semiclassical Gross-Pitaevskii description are
insufficient for capturing the complicated dynamics of the BJJ, and a
many-body approach is necessary \cite{Sakmann09}. It is because
a time-evolving system will always create incoherent excitations above the BEC (which we will call quasiparticle (QP) excitations), even if it was initially prepared with all particles in the BEC, provided the frequencies characterizing its time dependence are larger than the lowest QP excitation energy, characterized by the trap frequency. If the BEC is initially prepared in an excited state, for instance, by a spatial modulation of the BEC amplitude or by an occupation imbalance of a BEC in a double-well potential, QP excitations out of the condensate are necessary to thermalize the system, i.e., to reach a thermal distribution of all states and to accommodate the entropy of the thermal state. Hence, one may expect that even in a non-integrable system one can prevent thermalization by making the trap small enough, so that the trap frequency exceeds the characteristic frequency of the time evolution. Multiple undamped (non-thermalizing) \cite{Albiez05,Kinoshita06} as well as strongly damped (thermalizing) \cite{Thywissen11} Josephson oscillations of BECs in double-well potentials have been observed experimentally in traps with different sizes and particle numbers. The above discussion exemplifies the potential necessity of considering incoherent excitations in any temporal evolution of a BEC.

A first, general description of Bose many-body systems beyond the Gross-Pitaevskii means field dynamics was developed in the seminal works by
Beliaev \cite{Beliaev58},  Kadanoff and Baym \cite{Kadanoffbook},  Kane and Kadanoff \cite{Kane}, and  Hohenberg and Martin \cite{Hohenberg}. A kinetic theory which includes collisions between then  condensate and non-condensate part of the system was developed by Griffin, Nikuni and Zaremba \cite{Griffinbook}. Further approaches to a quantum kinetic description can be found in  Refs. \cite{Stoof01,Baier,Rey04} and references therein.  However, the wide applicability of these  methods is somewhat hampered by the large number of variables inherent to the spatial representation, and/or (as well as in numerical methods \cite{Sakmann09}) suitability for only moderate number of atoms.  

In the present paper we lay out a general, tractable method for the quantum
dynamics of Bose-Einstein condensed systems in the presence of non-condensate
excitations. It is based on a systematical expansion of the bosonic field
operator in the eigenstates of a non-interacting Bose system in an arbitrary
trap potential \cite{Mauro09}, where the operators for the lowest-lying states
are replaced by their respective, time-dependent condensate amplitudes and the full 
quantum dynamics of the excited states is retained. We derive the coupled 
equations of motion for the classical condensate propagator and the quantum
propagator of the excitations. By using this eigenstate representation, 
the spatial dependence is cast into the eigenstate wavefunctions and the 
number of degrees of freedom can be limited to the number of
trap states to be considered, thus keeping the quantum problem tractable even
for complicated trap potentials,  depending only on a few system parameters. 
This allows us to treat any number of particles. 

While our formalism is completely general, in this work we apply it to a system of an interacting Bose gas
confined in a double-well potential with an initial population imbalance of the
BECs in the left and in the right potential well. 
This system is established to exhibit Josephson oscillations \cite{Smerzi97,Albiez05, Levy07}.

The Josephson effect, well known from superconductors
\cite{Josephson62,Barone} and superfluids  \cite{Pereverzev,Sukhatme} has its
perculiarities in  BEC systems
\cite{Javanainen86,Jack96,Milburn97,Smerzi97,Radzihovsky10,Gati07}. Apart from
already mentioned experiments it has been also observed in 1D Bose Josephson
junctions on a chip \cite{Betz11} and arrays of Bose-Einstein
condensates \cite{Cataliotti01}. The BJJ is usually described in
terms of the oscillating condensate population imbalance
\beq
z(t)=\frac{N_1(t)-N_2(t)}{N_1(0)+N_2(0)},
\label{def_imbalance}
\eq 
where $N_{1}$ ($N_2$) is the number of particles in the left (right)
well. The dynamics of $z(t)$ is very  
 sensitive to interactions. One distinguishes two regimes: a delocalized
 regime with the time average $\langle z(t) \rangle =0$, and a self-trapped
 (ST) regime with $\langle z(t) \rangle \neq 0$  \cite{Smerzi97,Albiez05}. The
 transition from the delocalized to the self-trapped regime
 occurs when the initial population imbalance $z(0)$ exceeds a critical value $z_c(0)$ for fixed interaction, or equivalently, 
the interaction strength between bosons becomes greater than a
certain critical value, which depends on $z(0)$.  

Once incoherent single-particle excitations are taken into account, the
semiclassical dynamics of the BJJ drastically changes
\cite{Mauro09,Zapata98,Zapata03,Sakmann09}.  First of all, the self-trapped
state is destroyed by qp excitations
\cite{Mauro09,Sakmann09}.
Secondly, we have shown that the non-equilibrium QP
  dynamics does not set in instantaneously
 as the Josephson coupling is switched on, but with some delay, described by a
 time scale $\tau_c$ \cite{Mauro09}. For $t>\tau_c$  QPs are
 excited in an avalanche manner leading to fast 
Rabi-like oscillations of the 
QP occupation numbers and of the condensates between the wells. We provide a
detailed analysis of the origin of $\tau_c$ and its dependence on systems parameters. 

The paper is organized as follows.
In Section \ref{model} we define the model Hamiltonian and main parameters of the problem. In Section \ref{formalism} we present the general formalism in terms of Keldysh Dyson equations for condensed and non-condensed particles. In Section \ref{Hartree-Fock} we apply our formalism to a double-well potential and consider the effect of QPs on the BJJ within Bogoliubov-Hartree-Fock approximation. We discuss our results for particle imbalances, and inverse $\tau_c$ versus various system parameters in Section \ref{Results}. Main conclusions and perspectives are given in Section \ref{Conclusions}.

\section{The model Hamiltonian}
 \label{model}

A system of weakly interacting bosons, trapped in an external double-well potential (Fig.~1), is most generally described by the Hamiltonian  
\beq
H=\int d {\bf r} \hat \Psi^{\dg}({\bf r},t)\left(-\frac{\nabla^2}{2m}+V_{ext}({\bf
    r},t)\right)\hat \Psi({\bf r},t)
+\frac{g}{2}\int d {\bf r} \hat \Psi^{\dg}({\bf r},t)\hat \Psi^{\dg}({\bf
  r},t)\hat \Psi({\bf r},t)\hat \Psi({\bf r},t),
\label{gen_ham}
\eq
where $\hat \Psi({\bf r},t)$ is a bosonic field operator, and we assumed a contact interaction between the bosons with $g=4\pi a_s/m$
($a_s$ is the s-wave scattering length). $V_{ext}$ is the 
external double-well trapping
potential. 

\begin{figure}[!bt]
\begin{center}
\includegraphics[width=0.45\textwidth]{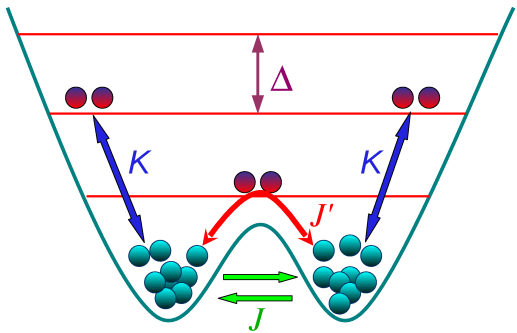}\vspace*{-0.8em} 
\end{center}
\caption{Bose-Einstein condensate in a double well potential with discrete energy levels. Blue particles are condensed ones, and red particles are the particles excited out of the condensates. $J$ is the Josephson coupling between the wells, $J'$ is the QP-assisted Josephson coupling, $K$ is the interaction between the QPs and the condensed particles, $\Delta$ is the interlevel splitting. }
\label{setup}
\end{figure} 

We now expand the field operator $\hat \Psi({\bf r},t)$ in terms of a complete basis  $\mathds{B}=\{\varphi_-,\varphi_+,\varphi_1,\varphi_2, \dots \varphi_M  \}$  of the exact single-particle eigenstates of the double well potential $V_{ext}({\bf r}, t>0)$ after the coupling between the wells is turned on. 
Note that the ground state wavefunction has a zero in the barrier between the wells, thus minimizing its energy, i.e., for a symmetric double-well, it
is parity antisymmetric, while the first excited state is symmetric. 
Hence, we denote the ground state wavefunction of the double well by 
$\varphi_-$, the first excited state wavefunction by $\varphi_+$, the 
second excited state by $\varphi_1$ and so on. 
The field operator reads in the eigenbasis of the double-well potential,
\beq
\hat \Psi({\bf r},t)=\phi_1({\bf r})\hat b_{01}(t)+\phi_2({\bf
  r})\hat b_{02}(t)+\sum_{n=1}^{M}\varphi_n({\bf r}) \hat b_n(t)\ ,
\label{field_operator1}
\eq
where we have applied the transformation, 
$\hat b_{01} (t) = (\hat b_- + \hat b_+)/\sqrt{2}$,  
$\hat b_{02} (t) = (\hat b_- - \hat b_+)/\sqrt{2}$ on the operators 
$\hat b_{\pm}(t)$ for particles in the $\varphi _{\pm}$ subspace, 
with the wavefunctions 
$\phi_1({\bf r})=(\varphi_-({\bf r})+\varphi_+({\bf r}))/\sqrt{2}$ and
$\phi_2({\bf r})=(\varphi_-({\bf r})-\varphi_+({\bf r}))/\sqrt{2}$.
Since the $\varphi_+({\bf r})$ ($\varphi_-({\bf r})$) have the same 
(the opposite) sign in the two wells, the $\phi_{1,2}({\bf r})$ are 
localized in the left or right well, respectively, i.e., they 
approximately constitute the ground state wavefunctions of 
the left and right well \cite{footnote}. We now perform the 
Bogoliubov substitution for the (approximate) ground states in each of
the two wells, 
\beq
\hat b_{0\alpha}(t) \to a_{\alpha}(t) = 
\sqrt{N_{\alpha}(t)}e^{i\theta_{\alpha}(t)},
\label{cond_wf}
\eq
$\alpha=1,2$, where $N_{\alpha}$ and $\theta_{\alpha}$  are the
number of particles and the phase of the condensate in the left (right) well
of the potential. The field operator then reads,
\beq
\hat \Psi({\bf r},t)=\phi_1({\bf r}) a_1(t)+\phi_2({\bf
  r}) a_2(t)+\sum_{n=1}^{M}\varphi_n({\bf r}) \hat b_n(t)\ ,
\label{field_operator2}
\eq
The first two terms in Eq. (\ref{field_operator2}) constitute the usual 
two-mode approximation for a condensate in a double well 
\cite{Milburn97,Smerzi97,Gati07}. The Bogoliubov substitution neglects phase 
fluctuations in the ground states of each of the potential wells, while
the full quantum dynamics is taken into account for the excited states, 
$\varphi_n$, $n=1,\ 2,\dots,\ M$.
This is justified when the BEC particle 
numbers are sufficiently large, $N_{\alpha}\gg 1$, e.g., 
for the experiments \cite{Albiez05}. The applicability of the semiclassical 
approximation has been discussed in detail in 
Refs.~\cite{Zapata03,Zapata98,Pitaevskii01} and has been tested 
experimentally in Ref.~\cite{Esteve08}. 
Note that Eq.~(\ref{field_operator2}) is a complete expansion in the 
single-particle eigenbasis of the double well. Therefore, one may ascribe
to  the low-lying excitations above the states $\phi_1$, $\phi_2$ the 
character of collective BEC excitations, while the high-energy excitation
have the character of single-particle excitations, as is well-known for 
the case of a translationally invariant BEC \cite{Stoof}.  
In the numerical evaluations we will limit the number of levels which
can be occupied by the QPs to $M=5$.

At time $t=0$ the barrier between the wells is suddenly lowered, and a Josephson link is established between the two condensate reservoirs. The Hamiltonian  for our system for $t>0$ is then 
\beq
H=H_{BEC}+H_{qp}+H_{mix}.
\label{full_ham}
\eq
$H_{BEC}$ describes only condensate particles
\bea
H_{BEC}=E_0\sum_{\alpha=1}^2a_{\alpha}^*a_\alpha+\frac{U}{2}\sum_{\alpha=1}^2a_{\alpha}^*a_{\alpha}^*a_\alpha a_\alpha 
-J(a_1^*a_2+a_2^*a_1).
\label{H_BEC}
\ea
Here we consider symmetric wells with equal interparticle  interactions:
\bea
E_0&=&\int d{\bf r}\left[ \frac{\hbar^2}{2m}|\nabla \phi_{1,2}({\bf r})|^2+\phi_{1,2}^2 V_{ext}({\bf r}) \right], \\
U&=&g\int d{\bf r} \, |\phi_{1,2}({\bf r})|^4,
\ea
and the Josephson coupling
\beq
J=-2\int d {\bf r} \left [ \frac{\hbar^2}{2m}(\nabla \phi_1\nabla \phi_2)+\phi_1 \phi_2 V_{ext}({\bf r}) \right ].
\eq
Note that there is no time dependence of $V_{ext}$ for the times consisdered, 
$t>0$, since the Josephson tunnelling $J$ was turned on for times $t >0$, and kept constant afterwards. 
$H_{qp}$ corresponds to the single-particle excitations to higher levels
\beq
H_{qp}=\sum_{n=1}^M \epsilon_n\hat b_n^{\dagger} \hat b_n+\frac{U'}{2}\sum_{n,m=1}^M\sum_{l,s=1}^M \hat b_m^{\dagger}  \hat b_n^{\dagger} \hat b_l \hat b_s,
\eq
where 
\beq
\epsilon_n=\int d{\bf r} \varphi_n({\bf r})\left(
-\frac{\nabla^2}{2m}+V_{ext}({\bf r})\right) \varphi_n({\bf r}) \ .
\label{epsilon}
\eq
For computational simplicity we assume in the following that the levels are
equidistant, $\epsilon_n=n\Delta$, with $\Delta$ the level spacing. 
This assumption does not restrict the generality of the formalism. 
It becomes exact, if $V_{ext}({\bf r})$ 
is harmonic for large energies, i.e., for $V_{ext}({\bf r}) \gtrsim \epsilon_1$. 

$U'$ is the repulsive interaction between single-particle excitations
\beq
U'=g\int d{\bf r} \varphi_n({\bf r})\varphi_m({\bf r})\varphi_l({\bf r})\varphi_s({\bf r}). 
\label{int_Up}
\eq

Finally, mixing between the BECs and the system of QP excitations is described by
\beq
H_{mix}=\frac{1}{2}\sum_{\alpha,\beta=1}^2\sum_{n,m=1}^MK_{\alpha \beta nm}\large[(a_{\alpha}^*a_{\beta}^*\hat b_n\hat b_m+h.c.) + 4a_{\alpha}^*a_{\beta}\hat b_n^{\dagger}\hat b_m  \large].
\eq
Here 
\beq
K_{\alpha \beta n m }=g\int d{\bf r} \phi_{\alpha}({\bf r})\phi_{\beta}({\bf r})\varphi_n({\bf r})\varphi_m({\bf r}).
\label{int_K}
\eq
It is convenient to introduce two interaction constants
\bea
K&=&2K_{11nm}=2K_{22nm} \\
J'&=&2K_{12nm}=2K_{21nm},
\ea
so that terms proportional to $J'$  describe QP-assisted Josephson 
tunneling. $H_{mix}$ is then
\bea
H_{mix}&=&J'\sum_{n,m=1}^M \left[ (a_1^*a_2+a_2^*a_1)\hat b_n^{\dg}\hat b_m+\frac{1}{2}(a_1^*a_2^*\hat b_n\hat b_m +h.c.) \right]\nn \\
&+&K\sum_{\alpha=1}^2\sum_{n,m=1}^M \left[a_{\alpha}^*a_{\alpha}\hat b_n^{\dg}\hat b_m+\frac{1}{4}(a_{\alpha}^*a_{\alpha}^*\hat b_n \hat b_m+h.c.)\right].
\ea
Hereafter all our energy scales are counted from $E_0$. The main parameters in
our model are, hence, the interlevel spacing $\Delta$, the QP-assisted Josephson coupling $J'$ and the QP-BEC interaction $K$ (Fig. \ref{setup}).

In the next chapters we proceed to derive the equations of motion for the condensates and the QP excitations described by the quantum  operators $\hat b, \hat b^{\dg}$.

\section{General formalism}

\label{formalism}


\subsection{Quantum kinetic theory: Dyson equations and self-energies}

In this section we formulate the quantum kinetic theory of the Bose Josephson junction in terms of non-equilibrium Keldysh Green's functions \cite{Keldysh65} defined along the Keldysh time contour. 
To shorten the notations we denote the classical part of our field operator \eqref{field_operator2} as $\Psi_0({\bf r},t)$ 
and the quantum part as $\tilde \Psi({\bf r},t)$. 
We define correspondingly two  one-particle Green's functions, one for the QPs, ${\bf G}$, and one for the condensed particles, ${\bf C}$. 
\bea
{\bf G}(1,1' ) = -i \left(\begin{matrix} \la T_C \tilde \Psi(1) \tilde \Psi^\dg (1' ) \ra & \la T_C \tilde \Psi(1) \tilde \Psi (1' ) \ra\\
\la T_C \tilde \Psi^\dg (1) \tilde \Psi^\dg (1' ) \ra &\la T_C \tilde \Psi^\dg (1) \tilde \Psi (1') \ra  \end{matrix} \right)
= \left(\begin{matrix} G(1,1' ) & F(1,1' ) \\ \ov{F}(1,1' ) & \ov{G}(1,1' )   \end{matrix} \right),
\label{GF1}
\ea 
where $1\equiv ({\bf r},t)$, $1'\equiv ({\bf r}',t')$, and $T_C$ denotes time ordering along the Keldysh contour, i.e. each of the bosonic Green's functions $G$ and $F$ will be a $2\times2$ matrix in Keldysh space. The condensate propagator ${\bf C}$ is classical with trivial time ordering,
\beq
{\bf C}(1,1')=-i\left( \begin{matrix}\Psi_0 (1) \Psi_0 ^* (1') & \Psi_0 (1) \Psi_0 (1') \\ \Psi_0 ^* (1) \Psi_0 ^* (1') & \Psi_0 ^* (1) \Psi_0 (1')  \end{matrix}\right).
\label{cond1}
\eq
For the derivation of the equation of motion for ${\bf G}$ and ${\bf C}$, we
proceed in the standard way (see, for instance
Refs. \cite{Griffinbook,Rammerbook}). We first write down the associated
Dyson equations for both Green's functions on the Keldysh
contour, which read as follows,
\bea
\int_C d2 \left[{\bf G }_0 ^{-1} (1,2) - {\bf S}^{HF} (1,2) \right] {\bf C} (2, 1' )
 =  \int_C d2\,{\bf S }(1,2)  {\bf C} (2, 1' ),\label{dyson-cond-keldysh} 
\ea
\bea
\int_C d 2 \left[{\bf G }_0 ^{-1} (1,2) - {\bf \Sigma}^{HF} (1,2) \right] {\bf G} (2, 1' )
 = \mathds{1} \delta(1-1')+ \int_C d 2\, {\bf \Sigma } (1,2)  {\bf G} (2, 1' ) ,\label{dyson-noncond-keldysh}
\ea
where $\int_C d2 \equiv \int d {\bf r}_2 \int_C d t_2 $ and the subscript $C$
denotes the time integration along the Keldysh contour. For convenience, we have
decomposed the self-energies into the first-order (self-consistent Hartree-Fock) contribution,
${\bf S}^{HF}$ (${\bf \Sigma}^{HF}$) and the higher-order part, ${\bf S}$ (${\bf \Sigma}$), 
which accounts for inelastic collisions. Collisional self-energies, which are
responsible for dissipation and equilibration, can be derived within
self-consistent second-order approximation. 
The bare $2\times2$ propagator ${\bf G}_0$ is defined as
\beq
{\bf G }_0 ^{-1} (1,1') = \delta(1-1') \left[i \tau_3 \partiel{}{t_1} -\left(-\frac{1}{2m}\Delta_1 + V_{ext}(1) \right)\mathds{1} \right] 
\eq
where 
\beq
\tau_3 = \left(\begin{matrix} 1 & 0 \\ 0 & -1 \end{matrix}\right),\label{pauli3}
\eq
and $\mathds{1}$ is the $2\times2$ identity matrix in Bogoliubov space. 
The Hartree-Fock self-energies correspond to the one-particle irreducible diagrammatic contributions of the first order in the interaction strength $g$
\bea
&{\bf S}^{HF} (1,1') = ig \delta(1-1') {\bf G}(1,1')
&+ \frac{i}{2}g \delta(1-1')\left\{\mathrm{Tr}\left[{\bf C}(1,1)\right] +\mathrm{Tr}{\left[ {\bf G}(1,1)\right]}  \right\}\mathds{1},
\label{HF-cond-self} 
\ea
\bea
{\bf \Sigma}^{HF} (1,1') &=& ig \delta(1-1') \left\{{\bf G}(1,1') +{\bf C}(1,1')\right\}  \nn \\
&+& \frac{i}{2}g \delta(1-1')\left\{\mathrm{Tr}\left[{\bf C}(1,1)\right] +\mathrm{Tr}{\left[ {\bf G}(1,1)\right]}  \right\}\mathds{1}.
\label{HF-noncond-self}
\ea
Parametrizing the Keldysh contour in terms of real time variable we obtain from Eqs. (\ref{dyson-cond-keldysh}) and (\ref{dyson-noncond-keldysh}), 
\bea
\int\limits_{-\infty}^{\infty} d2 \left[{\bf G }_0 ^{-1} (1,2) - {\bf S}^{HF} (1,2) \right] {\bf C} (2, 1' )
 = -i \int\limits_{-\infty}^{t_1} d2  {\bm\gamma } (1,2){\bf C} (2, 1' ),\label{dyson-cond-realtime} 
\ea
\bea
\int\limits_{-\infty}^{\infty} d 2 \left[{\bf G }_0 ^{-1} (1,2) - {\bf \Sigma}^{HF} (1,2) \right] {\bf G}^\gtrless (2, 1' )= \nn \\ -i 
\left[\int\limits_{-\infty}^{t_1} d 2\, {\bf \Gamma } (1,2)  {\bf G}^\gtrless (2, 1' ) - \int\limits_{-\infty}^{t_{1'}}d 2 {\bf \Sigma}^\gtrless (1,2) {\bf A}(2,1')\right].
\label{dyson-noncond-realtime}
\ea
Here we introduced the QP spectral function
\beq
{\bf A}(1,1') = i\left[{\bf G}^> (1,1') -{\bf G}^< (1,1')\right] 
\label{eq:spectral}
\eq
and analogously,
\beq
{\bf \Gamma}(1,1') = i\left[{\bf \Sigma}^> (1,1') -{\bf \Sigma}^< (1,1')\right],
\eq
and
\beq
{\bm \gamma}(1,1') = i\left[{\bf S}^> (1,1') -{\bf S}^< (1,1')\right]
\eq
for the corresponding self-energies not containing the Hartree-Fock contributions.  The "lesser" and "greater" Green's functions are defined as usual \cite{Rammerbook},
\bea
{\bf G}^<(1,1' )= -i \left(\begin{matrix} \la \tilde \Psi^\dg(1') \tilde \Psi (1 ) \ra & \la \tilde \Psi(1') \tilde \Psi (1 ) \ra\\
\la \tilde \Psi^\dg (1') \tilde \Psi^\dg (1 ) \ra &\la \tilde \Psi (1') \tilde \Psi^\dg (1)  \ra \end{matrix} \right) 
=\left(\begin{matrix} G^<(1,1' ) & F^<(1,1' ) \\ \ov{F}^<(1,1' ) & \ov{G}^<(1,1' )   \end{matrix} \right), \\
{\bf G}^>(1,1' )= -i \left(\begin{matrix} \la \tilde \Psi (1 ) \tilde \Psi^\dg(1')  \ra & \la  \tilde \Psi (1 ) \tilde \Psi(1')\ra\\
\la  \tilde \Psi^\dg (1 ) \tilde \Psi^\dg (1') \ra &\la  \tilde \Psi^\dg (1) \tilde \Psi (1')  \ra  \end{matrix} \right) 
=\left(\begin{matrix} G^>(1,1' ) & F^>(1,1' ) \\ \ov{F}^>(1,1' ) & \ov{G}^>(1,1' )   \end{matrix} \right) \ .
\label{lesser}
\ea 
It is, furthermore, convenient to introduce the so-called statistical propagator (which
is just  the Keldysh component of the QP Green's function 
${\bf G}^K(1,1')$ divided by two),
\beq
{\bf F}(1,1') = \frac{{\bf G}^> (1,1') +{\bf G}^< (1,1')}{2},
\label{eq:statistical}
\eq
and rewrite our equations in terms of the spectral function and the statistical propagator which contain the information about the spectrum and the occupation number, respectively. We hence get,
\bea
\int\limits_{-\infty}^{\infty} d 2 \left[{\bf G }_0 ^{-1} (1,2) - {\bf \Sigma}^{HF} (1,2) \right] {\bf A} (2, 1' )=
-i \int\limits_{t_{1'}}^{t_1} d 2\, {\bf \Gamma } (1,2)  {\bf A} (2, 1' ) 
 \label{dyson-spectral}
 \ea
and
\bea
\int\limits_{-\infty}^{\infty} &d2& \left[{\bf G }_0 ^{-1} (1,2) - {\bf \Sigma}^{HF} (1,2) \right] {\bf F} (2, 1' ) = \nn \\
 &-&i \left[\int\limits_{-\infty}^{t_1} d 2\, {\bf \Gamma } (1,2)  {\bf F} (2, 1' ) - \int\limits_{-\infty}^{t_{1'}}d 2 {\bf \Pi} (1,2) {\bf A}(2,1')\right],
 \label{dyson-statistical}
\ea
where ${\bm \Pi}$ is defined as,
\beq
{\bm \Pi}(1,1') = \frac{{\bf \Sigma}^> (1,1') +{\bf \Sigma}^< (1,1')}{2}.
\eq
Eqs.~(\ref{dyson-spectral}) and (\ref{dyson-statistical}) were obtained by taking the difference and the sum of Eq.~(\ref{dyson-noncond-realtime}) for the "greater" and "lesser" propagator, and then inserting the definition of the spectral function (\ref{eq:spectral}) and the statistical function (\ref{eq:statistical}), respectively. The Hartree-Fock self-energies in terms of the spectral and statistical functions read
 \bea
&{\bf S}^{HF} (1,1') = ig \delta(1-1') {\bf F}(1,1')
&+ \frac{i}{2}g \delta(1-1')\left\{\mathrm{Tr}\left[{\bf C}(1,1)\right] +\mathrm{Tr}{\left[ {\bf F}(1,1)\right]}  \right\}\mathds{1},
\ea
\bea
{\bf \Sigma}^{HF} (1,1')=ig \delta(1-1') \left\{{\bf F}(1,1')+{\bf C}(1,1')\right\}  \nn \\
+ \frac{i}{2}g \delta(1-1')\left\{\mathrm{Tr}\left[{\bf C}(1,1)\right]+\mathrm{Tr}{\left[ {\bf F}(1,1)\right]}\right\}\mathds{1},
\ea
where we used $\left. {\bf G}^< (1,1')\right |_{t_{1} =t_{1'}} =\left. {\bf F}(1,1')\right|_{t_{1} =t_{1'}}+i\mathds{1}\delta({\bf r}_1-{\bf r}_{1}')/2$. (The terms proportional to $\delta({\bf r}_1-{\bf r}_{1}')$ are not explicitly written in the equations to shorten the notation). 
It is worth mentioning that these self-energies only contain the symmetrized  two-point propagators instead of the time-ordered ones. The non-condensate and the anomalous density are then expressed in terms of the propagators $G^< $ and $F^< $,  evaluated at equal points in space and time.

We now derive from our general Dyson equations the equations of motion for the condensates and the QP excitations in the eigenbasis $\mathds{B}$ of the trap
potential. 


\subsection{Condensate equation of motion in the trap eigenbasis}
\label{condensates}
Using the bosonic field operator (\ref{field_operator2}), we straightforwardly obtain for the propagators \eqref{GF1} and \eqref{cond1}, the spectral function (\ref{eq:spectral}) and the statistical function (\ref{eq:statistical}) the following expressions,
\beq
{\bf G}(1,1' )=\sum_{n,m=1}^M  \varphi_m ({\bf r}) \varphi_n ({\bf r}') {\bf G}_{nm}(t, t')
\eq
\beq
{\bf C} (1,1')  =  \sum_{\alpha,\beta =1} ^2 \phi_\alpha ({\bf r}) \phi_\beta ({\bf r}') {\bf C}_{\alpha \beta}(t, t'),\label{cond2}
\eq
\beq
{\bf A} (1,1')  =  \sum_{n,m=1}^M  \varphi_m ({\bf r}) \varphi_n ({\bf r}') {\bf A}_{nm}(t, t'),\label{spectral1}
\eq
\beq
{\bf F} (1,1')  =  \sum_{n,m=1}^M  \varphi_m ({\bf r}) \varphi_n ({\bf r}') {\bf F}_{nm}(t, t'),\label{statistical1}
\eq
where 
\bea
{\bf G}_{nm}(t,t')&=&-i\left( \begin{matrix}
\la T_C\hat b_n(t)\hat b_m^\dg (t') \ra & \la T_C\hat b_n(t)\hat b_m (t') \ra \\
\la T_C\hat b_n^\dg(t)\hat b_m^\dg (t') \ra  & \la T_C\hat b_n^\dg(t)\hat b_m (t') \ra
\end{matrix}
\right) 
=\left(\begin{matrix} G_{nm}(1,1' ) & F_{nm}(1,1' ) \\ \ov{F}_{nm}(1,1' ) & \ov{G}_{nm}(1,1' )   \end{matrix} \right) \\
{\bf C}_{\alpha\beta}(t,t')&=&-i\left( \begin{matrix}
a_\alpha(t)a_\beta^*(t') & a_\alpha(t) a_\beta(t') \\
a_\alpha^*(t)a_\beta^*(t')  & a_\alpha^*(t) a_\beta(t')
\end{matrix},
\right) ,
\label{cond}\\
{\bf A}_{nm}(t,t')&=&\left(\begin{array}{cc}
\text{A}^G _{nm}(t,t')& \text{A}^F _{nm}(t,t') \\
-\text{A}^F _{nm}(t,t')^* & -\text{A}^{G} _{nm}(t,t')^*
\end{array}
\right) , \label{spectral2}\\
{\bf F}_{nm}(t,t')&=&\left(\begin{array}{cc}
\text{F}^G _{nm}(t,t')& \text{F}^F _{nm}(t,t') \\
-\text{F}^F _{nm}(t,t')^* & -\text{F}^{G} _{nm}(t,t')^*\end{array}
\right),
\label{statistical2}
\ea
with $\text{A}^G _{nm}= i(G^>  _{nm}- G^< _{nm})$, $\text{A}^F _{nm}= i(F^>  _{nm}- F^< _{nm})$ and $\text{F}^G _{nm}= (G^>  _{nm}+G^<  _{nm} )/2$, and $\text{F}^F _{nm}= (F^>  _{nm}+F^<  _{nm} )/2$.
Here and in the following, Greek indeces, $\alpha,\,\beta = 1,\,2$, 
refer to the condensates in the left and right wells, and latin indeces,
$n,\,m =1,\,2,\,\dots,\,M$, denote the QP levels. Moreover, we will adopt the
sum convention, i.e., Greek or Latin indeces appearing in a term twice are 
summed over.

\begin{figure}[!bt]
\begin{center}
\includegraphics[width=0.45\textwidth]{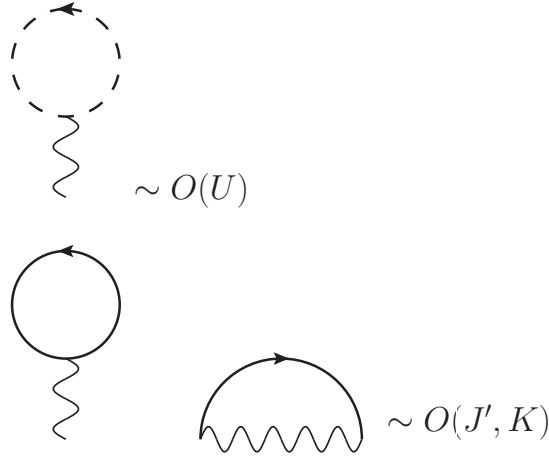}\vspace*{-0.8em} 
\end{center}
\caption{Hartree-Fock self-energies for the condensate propagator ${\bm
    S}^{HF} _{\alpha\beta}$. The solid and dash lines represent the $2\times2$
  single particle excitations propagator and the condensate propagator,
  respectively. The wavy lines denote the interactions $K, J'$ or
  $U$, depending on which particles are involved in the process. 
Note that the condensate and the non-condensate propagators 
in these expressions are non-diagonal in the condensate index $\alpha$ and 
the QP index $m$, respectively, since they are the interacting propagators
calculated self-consistently from the equations of motion 
(\ref{EOMcond}), (\ref{evolution-spectral1}), (\ref{evolution-statistical1}).
}
\label{fig:selfenergy-cond-hf}
\end{figure} 
Inserting Eqs. (\ref{cond2}), (\ref{spectral1}) and (\ref{statistical1}) into
the Dyson equation for the condensate Green's function,
(\ref{dyson-cond-realtime}) we obtain the Dyson equation for ${\bf
  C}_{\alpha\beta}(t,t')$, where the position dependence is absorbed
in the parameters $E_0,U,U',K,J$ and $J'$,
\bea
 \int\limits_{-\infty}^\infty d \ov{t} \left[{\bf G }_{0,\alpha\gamma} ^{-1} (t,\ov{t}) - {\bf S}^{HF} _{\alpha\gamma}(t,\ov{t}) \right] {\bf C}_{\gamma\beta}(\ov{t} , t'  ) 
=-i \int\limits_{-\infty}^{t} d \ov{t}  {\bm \gamma}_{\alpha\gamma}(t,\ov{t}){\bf C}_{\gamma\beta} (\ov{t} , t').\label{dyson-cond-eigen}  
\ea
In this equation, ${\bm \gamma}_{\alpha\beta} = {\bm S}^> _{\alpha\beta} - {\bm S}^< _{\alpha\beta}$ accounts for collisions.
The bare propagator is given by
\beq
{\bf G }_{0,\alpha\beta} ^{-1} (t,t') = \left[i \tau_3 \delta_{\alpha\beta}\partiel{}{t} - \mathds{1} E_{\alpha\beta}\right] \delta(t -t') \ ,
\eq
with $E_{11}=E_{22}=E_0$ and $E_{12}=E_{21}=-J$.
The Hartree-Fock self-energies read,
\bea
{\bf S}^{HF} _{\alpha\alpha} (t,t') =\frac{i}{2}U\;\mathrm{Tr}\left[{\bf C}_{\alpha\alpha}(t,t)\right]\mathds{1}\delta(t-t')+ \nn \\ i\frac{K}{2}\sum_{n,m=1}^M \left\{\frac{1}{2} \mathrm{Tr}\left[{\bf F}^< _{nm}(t,t)\right]\mathds{1} + {\bf F}^< _{nm}(t,t)\right\}\delta(t-t') \\
{\bf S}^{HF} _{12} (t,t') ={\bf S}^{HF} _{21} (t,t')= \quad \quad \quad \quad \quad \quad \nn \\ i\frac{J'}{2}\sum_{n,m=1}^M \left\{\frac{1}{2} \mathrm{Tr}\left[{\bf F}^< _{nm}(t,t)\right]\mathds{1} + {\bf F}^< _{nm}(t,t)\right\}\delta(t-t'), 
\label{selfenergy-cond-hf}
\ea
In Fig. \ref{fig:selfenergy-cond-hf} the diagrammatic representation of the self-energy (\ref{selfenergy-cond-hf}) is given. 
After evaluating the integral on the left-hand side  of Eq. \eqref{dyson-cond-eigen}, the equation of motion for the condensate propagator reads,
\bea
\left[i \tau_3 \delta_{\alpha\gamma}\partiel{}{t} - \mathds{1} E_{\alpha\gamma} - {\bf S}^{HF} _{\alpha\gamma} (t)\right] {\bf C}_{\gamma\beta}(t , t'  ) 
=-i \int\limits_{-\infty}^{t} d \ov{t}  {\bm \gamma}_{\alpha\gamma}(t,\ov{t}){\bf C}_{\gamma\beta} (\ov{t} , t'),
\label{EOMcond}
\ea
where we used that ${\bf S}^{HF} _{\alpha\gamma} (t,t') = {\bf S}^{HF} _{\alpha\gamma} (t)\delta(t-t')$. The equation of motion for the time-dependent condensate amplitude $a_\alpha (t)$ can be obtained by taking the upper left component of Eq. \eqref{EOMcond} and then dividing by $a_\beta ^*(t')$,
\bea
i\partiel{}{t}a_\alpha=\left[ E_{\alpha\gamma}+S^{HF}_{\alpha\gamma} (t) \right]a_\gamma (t)+W^{HF} _{\alpha\gamma}(t)a_\gamma ^*(t)
-i\int\limits_{-\infty}^{t}d \ov{t} \left[ \gamma^S _{\alpha\gamma} (t,\ov{t}) a_\gamma (\ov{t}) + \gamma^ W_{\alpha\gamma} (t,\ov{t}) a_\gamma (\ov{t}) a_\gamma ^* (\ov{t})   \right],\quad
\label{evolution-condensate}
\ea
where $S^{HF}_{\alpha\beta}$ and $W^{HF} _{\alpha\beta}$ are the upper left and the upper right components of ${\bf S}^{HF} _{\alpha\beta}$, defined in Eq. (\ref{selfenergy-cond-hf}), i.e.,
\bea
{\bf S}_{\alpha\beta}^{HF}(t)=\left(\begin{matrix} S_{\alpha\beta}^{HF}(t) & W_{\alpha\beta}^{HF}(t) \\
W_{\alpha\beta}^{HF}(t)^* & S_{\alpha\beta}^{HF}(t)^* \end{matrix} \right).
\ea

 Similarly, $\gamma^S _{\alpha\beta}$ and $\gamma^
 W_{\alpha\beta}$ are the upper left and the upper right 
components of ${\bm\gamma}_{\alpha\beta}$, respectively.


\subsection{Quasiparticle equations of motion in the trap eigenbasis}
\label{quasiparticles}
We follow here the procedure analogous to the one described in
the previous section for the evolution Eqs. (\ref{dyson-spectral}) and
({\ref{dyson-statistical}) for non-condensate particle spectral and statistical functions. Inserting Eqs. (\ref{spectral1}) and (\ref{statistical1}) into Eqs. (\ref{dyson-spectral}) and ({\ref{dyson-statistical}) we get,
\bea
\int\limits_{-\infty}^\infty d \ov{t} \left[{\bf G}_{0,n\ell}^{-1}(t , \ov{t}) -{\bf \Sigma}_{n\ell}^{HF} (t , \ov{t}) \right] {\bf A}_{\ell m} (\ov{t} , t')
=-i\int\limits_{t'}^{t} d \ov{t}\, {\bf \Gamma}_{n\ell} (t, \ov{t}) {\bf A}_{\ell m} (\ov{t} , t')
\label{evolution-spectral}
\ea
\bea
\int\limits_{-\infty}^\infty d \ov{t} \left[{\bf G}_{0,n\ell}^{-1}(t , \ov{t}) -{\bf \Sigma}_{n\ell}^{HF} (t , \ov{t}) \right] {\bf F}_{\ell m} (\ov{t} , t')
=-i\Big{[}\int\limits_{-\infty}^{t} d \ov{t}\, {\bf \Gamma}_{n\ell} (t, \ov{t}) {\bf F}_{\ell m} (\ov{t} , t')\nn \\
\quad\quad-\int\limits_{-\infty}^{t'} d \ov{t}\, {\bf \Pi}_{n\ell}(t, \ov{t}) {\bf A}_{\ell m} (\ov{t} , t')\Big{]},
\label{evolution-statistical}
\ea
where ${\bm \Pi}_{nm} = ({\bm \Sigma}_{nm}^> + {\bm \Sigma}_{nm}^< )/2$ and
${\bf \Gamma}_{nm} = {\bf \Sigma}_{nm}^> - {\bf \Sigma}_{nm}^<$.
The bare one-particle propagator and the Hartree-Fock self-energy are given by
\beq
{\bf G}_{0,nm}^{-1}(t , t') = \left[i\tau_3 \partiel{}{t} - \epsilon_{n}\mathds{1} \right]\delta_{nm}\delta(t-t'),
\eq
\bea
&&{\bf \Sigma}_{nm}^{HF} (t ,t') = i \Big{(}\sum_{\alpha,\beta =1}^2 K_{\alpha\beta nm} \left\{{\bf C}_{\alpha\beta}(t,t)+\frac{1}{2}\mathrm{Tr}\left[{\bf C}_{\alpha\beta}(t,t)\right]\mathds{1}\right\}\nn\\
&& + U'\sum_{\ell,s\ne0} \left\{{\bf F}_{\ell s}(t,t)+\frac{1}{2}\mathrm{Tr}\left[{\bf F}_{\ell s}(t,t)\right]\mathds{1}\right\}\Big{)}\delta(t-t')
\label{selfenergy-non-cond-hf}\ ,
\ea
where $\epsilon_n$,  $U'$ and $K_{\alpha\beta nm}$ are defined in Eqs. \eqref{epsilon},  \eqref{int_Up} and \eqref{int_K}.
The first-order contributions to the self-energy of QP excitations in Eq. (\ref{selfenergy-non-cond-hf}) are shown in Fig. \ref{fig:selfenergy-non-cond-hf}.
\begin{figure}[!bth]
\begin{center}
\includegraphics[width=0.45\textwidth]{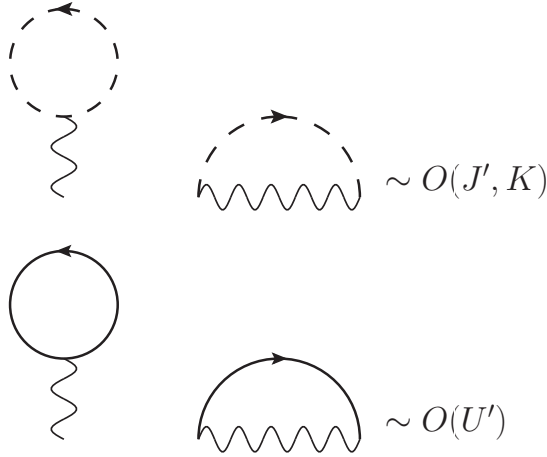}\vspace*{-0.8em} 
\end{center}
\caption{Diagrammatic contributions to the Hartree-Fock
  self-energies of non-condensed particles ${\bm\Sigma}^{HF} _{nm}$.
The solid and dashed lines represent  
the $2\times2$ single-particle excitation propagator and the condensate 
propagator, respectively. 
The wavy lines denote the interactions $K, J'$ or $U'$, 
depending on  which physical process of the Hamiltonian is involved.}
\label{fig:selfenergy-non-cond-hf}
\end{figure} 
Evaluation of the integrals on the left-hand side of
Eqs.~(\ref{evolution-spectral}) and (\ref{evolution-statistical}) yields,
\bea
\left[i\tau_3 \delta_{n\ell}\partiel{}{t} - \epsilon_{n}\delta_{n\ell}\mathds{1}  - {\bf \Sigma}^{HF} _{n\ell}(t)\right]{\bf A}_{\ell m} (t , t')
=-i\int\limits_{t'}^{t} d \ov{t}\, {\bf \Gamma}_{n\ell} (t, \ov{t}) {\bf A}_{\ell m} (\ov{t} , t'),
\label{evolution-spectral1}
\ea
\bea
\left[i\tau_3 \delta_{n\ell}\partiel{}{t} - \epsilon_{n}\delta_{n\ell}\mathds{1}  - {\bf \Sigma}^{HF} _{n\ell}(t)\right]{\bf F}_{\ell m} (t , t')
=-i\Big{[}\int\limits_{-\infty}^{t} d \ov{t}\, {\bf \Gamma}_{n\ell} (t, \ov{t}) {\bf F}_{\ell m} (\ov{t} , t')\nn \\
\quad\quad-\int\limits_{-\infty}^{t'} d \ov{t}\, {\bf \Pi}_{n\ell}(t, \ov{t}) {\bf A}_{\ell m} (\ov{t} , t')\Big{]},
\label{evolution-statistical1}
\ea
where we used that ${\bf \Sigma}^{HF} _{nm} (t,t') = {\bf \Sigma}^{HF} _{nm} (t)\delta(t-t')$.

Computing the evolution of ${\bf F}_{nm}$ for equal time arguments requires
special attention. In order to do it properly, we take the sum and the
difference of  Eq. (\ref{evolution-statistical1})  and its hermitean
conjugate, and evaluate it at equal times (see details in the next
section). Note, that  the components of ${\bf A}_{nm}(t,t)$ are fixed for all
times due to the bosonic commutation relations. One must also obey particle
and energy conservation by applying conserving approximations} - for details seethe Appendix.

Eqs.~(\ref{EOMcond}), (\ref{evolution-spectral1}) and
(\ref{evolution-statistical1}) constitute the general equations of motion for
the condensate and the non-condensate (spectral and statistical) propagators,
respectively. They are coupled via the self-energies which are 
functions of these propagators and must be evaluated self-consistently in
order to obtain a conserving approximation (see Appendix). The time-dependent Hartee-Fock
self-energies describe the dynamical shift of the condensate and the
single-particle levels due to the dynamical change of their occupation numbers
and their interactions. The higher order interaction terms on the right-hand
side of the equations of motion describe inelastic QP collisions. They are,
in general, responsible for QP damping, damping of the
condensate oscillations and for thermalization. In the present paper, we are
interested in how the non-equilibrium condensate oscillations induce
QP excitations and how the latter act back on the condensate
dynamics. Hence, we do not consider QP collision effects here.
The results including collisions will be published elsewhere \cite{MAH}.


\section{Equations of motion in Bogoliubov-Hartree-Fock approximation}

\label{Hartree-Fock}

We now neglect collisional terms and treat the non-equilibrium Bose Josephson junction within Bogoluibov-Hartree-Fock approximation. From Eqs. \eqref{evolution-condensate} and \eqref{evolution-statistical1} we get,
\beq
i\partiel{}{t}a_\alpha = \left[ E_{\alpha\gamma} + S^{HF}_{\alpha\gamma} (t)  \right]a_\gamma (t) + W^{HF} _{\alpha\gamma} (t) a_\gamma ^*(t), 
\label{evolution-condensate-hf}
\eq
\beq
i\tau_3 \delta_{n\ell}\partiel{}{t}{\bf F}_{\ell m} (t , t') = \left[ \epsilon_{n}\delta_{n\ell}\mathds{1}  + {\bf \Sigma}^{HF} _{n\ell}(t)\right]{\bf F}_{\ell m} (t , t').
\label{evolution-statistical1-hf}
\eq
It is not necessary to consider the equations for the spectral function ${\bf A}_{nm}$, since the only non-condensate related quantities appearing in the self-energies ${\bf S}^{HF} _{\alpha\beta}$ and ${\bf \Sigma}^{HF} _{nm}$ are the components of the statistical function ${\bf F}_{nm}$. This implies that Eqs.~(\ref{evolution-condensate-hf}) and (\ref{evolution-statistical1-hf}) decouple completely from the equations for ${\bf A}_{nm}$. This will not be the case when collisions are taken into account. 
Taking the difference of Eq.~(\ref{evolution-statistical1-hf}) and its hermitean conjugate, we get for ots diagonal component,
\bea
i\left(\partiel{}{t}\text{F}^G _{nm}(t,t') + \partiel{}{t'} \text{F}^G _{nm}(t,t')  \right)
=\left[\epsilon_{n}\delta_{n\ell} + \Sigma^{HF} _{n\ell}(t) \right]\text{F}^G _{\ell m}(t,t') - \Omega^{HF}_{n\ell} (t) \text{F}^F _{\ell m}(t,t')^*-\nn \\
\text{F}^G _{n\ell}(t,t')\left[\epsilon_{m}\delta_{m\ell} + \Sigma^{HF} _{\ell m}(t') \right] - \text{F}^F _{n\ell}(t,t') \Omega^{HF}_{\ell m} (t') ^* 
\label{evolution-Fg-hf}
\ea
The sum of Eq.~(\ref{evolution-statistical1-hf}) and its hermitean conjugate gives for its upper right component,
\bea
i\left(\partiel{}{t}\text{F}^F _{nm}(t,t') + \partiel{}{t'} \text{F}^F _{nm}(t,t')  \right)
=\left[\epsilon_{n}\delta_{n\ell} + \Sigma^{HF} _{n\ell}(t) \right]\text{F}^F _{\ell m}(t,t') - \Omega^{HF}_{n\ell} (t) \text{F}^G _{\ell m}(t,t')^* \nn \\
+ \text{F}^F _{n\ell}(t,t')\left[\epsilon_{m}\delta_{m\ell} + \Sigma^{HF} _{\ell m}(t') \right] + \text{F}^G _{n\ell}(t,t') \Omega^{HF}_{\ell m} (t') ^*
\label{evolution-Ff-hf}
\ea
Here we adopted the notation, 
\bea
{\bf \Sigma}_{nm}^{HF}(t)=\left(\begin{matrix} \Sigma_{nm}^{HF}(t) & \Omega_{nm}^{HF}(t) \\
\Omega_{nm}^{HF}(t)^* & \Sigma_{nm}^{HF}(t)^* \end{matrix} \right).
\ea
The self-energies $\Sigma^{HF}_{nm} (t)$ and $\Omega^{HF}_{nm} (t)$ are given by,
\bea
\Sigma^{HF}_{nm} (t)&=&K(N_1(t)+N_2(t))+J' a_1^*(t)a_2(t) 
+J'a_2^*(t)a_1(t)+2iU'\sum_{s,\ell} \text{F}^G _{s\ell}(t,t),  \label{self_energy_1} \\
\Omega^{HF}_{nm} (t)&=&\frac{K}{2}\sum_{\alpha=1}^2a_\alpha(t) a_\alpha(t) 
+J'a_1(t)a_2(t)+iU'\sum_{s,\ell} \text{F}^F _{s\ell}(t,t), 
\label{self_energy_2}
\ea
where $N_\alpha = a_\alpha ^* a_\alpha$, $\alpha=1,2$.
Evaluation of eq. (\ref{evolution-Fg-hf}) and (\ref{evolution-Ff-hf}) at equal times gives,
\bea
i\partiel{}{t}\text{F}^G _{nm}(t,t)= \left[\epsilon_{n}\delta_{n\ell} + \Sigma^{HF} _{n\ell}(t) \right]\text{F}^G _{\ell m}(t,t) 
 - \text{F}^G _{n\ell}(t,t)\left[\epsilon_{m}\delta_{m\ell} + \Sigma^{HF} _{\ell m}(t) \right] \nn \\
- \Omega^{HF}_{n\ell} (t) \text{F}^F _{\ell m}(t,t)^* - \text{F}^F _{n\ell}(t,t) \Omega^{HF}_{\ell m} (t) ^*, 
\label{evolution-Fg-hf1}
\ea
\bea
i\partiel{}{t}\text{F}^F _{nm}(t,t) =\left[\epsilon_{n}\delta_{n\ell} + \Sigma^{HF} _{n\ell}(t) \right]\text{F}^F _{\ell m}(t,t) 
 + \text{F}^F _{n\ell}(t,t)\left[\epsilon_{m}\delta_{m\ell} + \Sigma^{HF} _{\ell m}(t) \right]\nn\\
 - \Omega^{HF}_{n\ell} (t) \text{F}^G _{\ell m}(t,t)^* + \text{F}^G _{n\ell}(t,t) \Omega^{HF}_{\ell m} (t) ^* .
\label{evolution-Ff-hf1}
\ea
Note, that in our previous work \cite{Mauro09} we considered a special case of the functions $\text{F}^G _{nm}$ and $\text{F}^F _{nm}$ (or their equivalents)  being diagonal in energy space, i.e. being proportional to $\delta_{nm}$. In this work we do not adhere to this approximation and solve Eqs. \eqref{evolution-Fg-hf1} and \eqref{evolution-Ff-hf1} for all $n$ and $m$.
Plugging the expressions for $S^{HF}_{\alpha\beta}$ and $W^{HF}_{\alpha\beta}$  into Eq. \eqref{evolution-condensate-hf} we obtain,
\bea
i\partiel{}{t}a_1 (t) = \left[E_0 + UN_1(t) + iK \sum_{n,m} \text{F}^G _{nm} (t,t)\right]a_1 (t)
-\left[J -iJ'\sum_{n,m} \text{F}^G _{nm} (t,t)\right]a_2 (t) \nn\\
 +i\left[\frac{K}{2} a_1 ^* (t) + \frac{J'}{2}a_2^* (t)\right] \sum_{n,m} \text{F}^F _{nm} (t,t) .
\label{hartree-fock-eq-cond}
\ea
The equation for $a_2 (t)$ is obtained from Eq.\eqref{hartree-fock-eq-cond} by  replacing $a_1$ by $a_2$ and vice versa.

\begin{figure}[!h]            
\begin{center}
\includegraphics[width=0.9\textwidth]{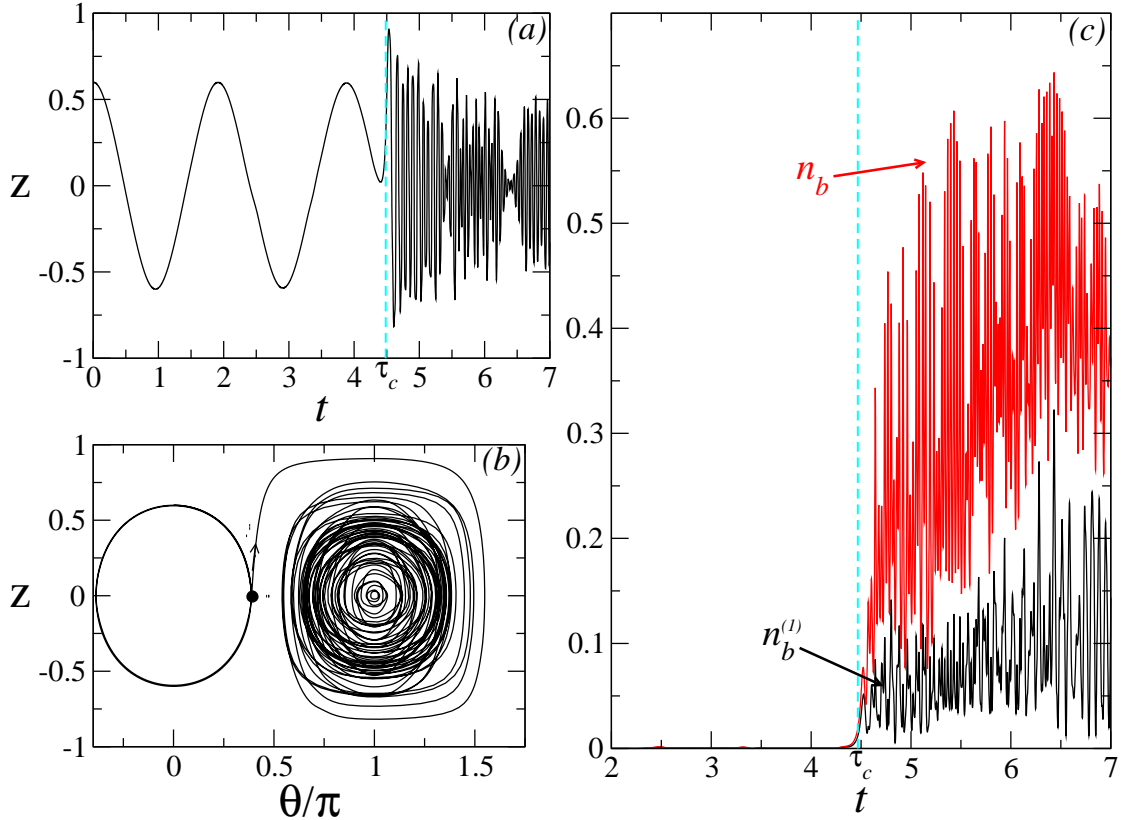}
\end{center}
\caption{
 ({\it a}) Time dependence of the particle imbalance $z$ for $z(0)=0.6, \theta(0)=0$, $N=500000$, $\Delta=20.6$, $u=u'=5$, $j'=60$.
({\it b}) Phase portrait of the junction. The thick point on the trajectory corresponds to $\tau_c$, i.e. to an establishing of non-equilibrium dynamics. The arrow shows direction $t>\tau_c$. ({\it c}) Time-dependence of the occupation by single-particle excited states of the first level $n_b^{(1)}$, the red curve is the sum of all five levels $n_b$.  The time is given in units of $1/J$. 
}
\label{imb_jos}
\end{figure} 

\begin{figure}[!h]
\begin{center}
\includegraphics[width=0.9\textwidth]{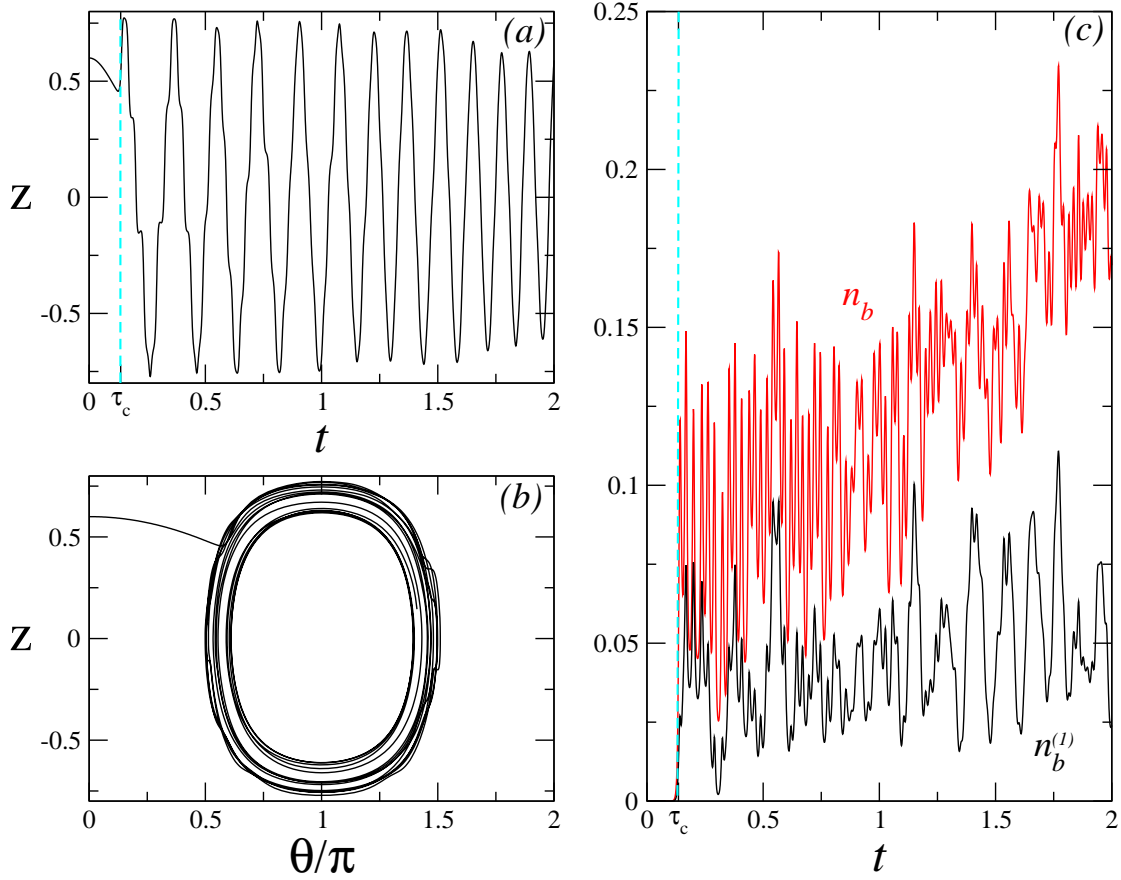}\vspace*{-0.5em}
\end{center}
\caption{
({\it a}) Time dependence of the particle imbalance $z$ for $z(0)=0.6, \theta(0)=0$, $N=500000$, $\Delta=25$, $u=u'=25$, $j'=60$.
({\it b}) Phase portrait of the junction.  ({\it c}) Time-dependence of the occupation by single-particle excited states of the first level $n_b^{(1)}$, the red curve is the sum of all five levels $n_b$. The time is given in units of $1/J$. 
 }
\label{imb_ST}
\end{figure}

\begin{figure}[!bt]
\begin{center}
\includegraphics[width=0.95\textwidth]{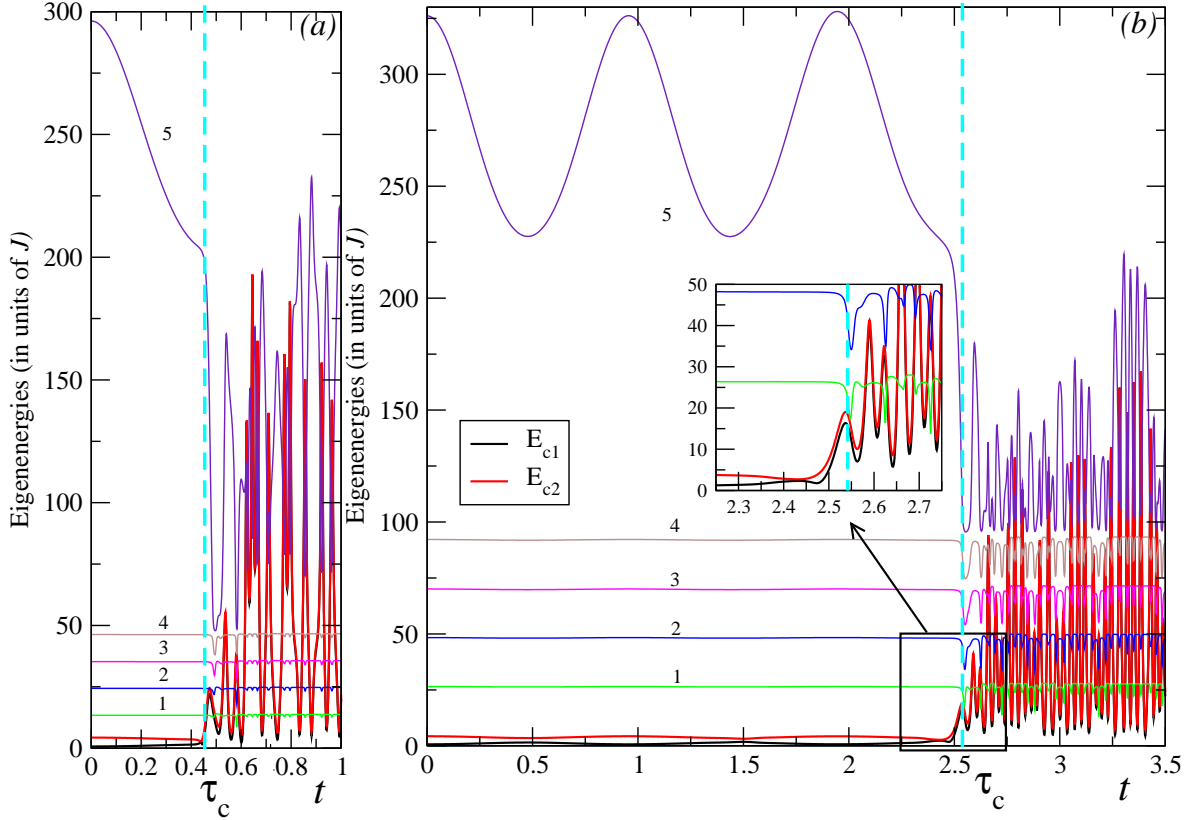}\vspace*{-0.5em}
\end{center}
\caption{
Instantaneous eigenenergies of the many-body Hamiltonian \eqref{full_ham} and extraction of $\tau_c$. Both plots are for $z(0)=0.6, \theta(0)=0$, total number of particles is $N=500000$, $u=u'=5$, $j'=60$ and bare level spacing  (a) $\Delta=10$ and (b) $\Delta=20$. The time is in  units of $1/J$. Curves $1$ to $5$ correspond to the QP eigenenergies $E_1,..,E_5$. The two  lowest curves in the plots correspond to eigenenergies of the condensate in the double well $E_{c1}$ and $E_{c2}$. A thick vertical line marks the characteristic time $\tau_c$. 
}
\label{levels}
\end{figure}

\begin{figure}[!bt]
\begin{center}
\includegraphics[width=0.95\textwidth]{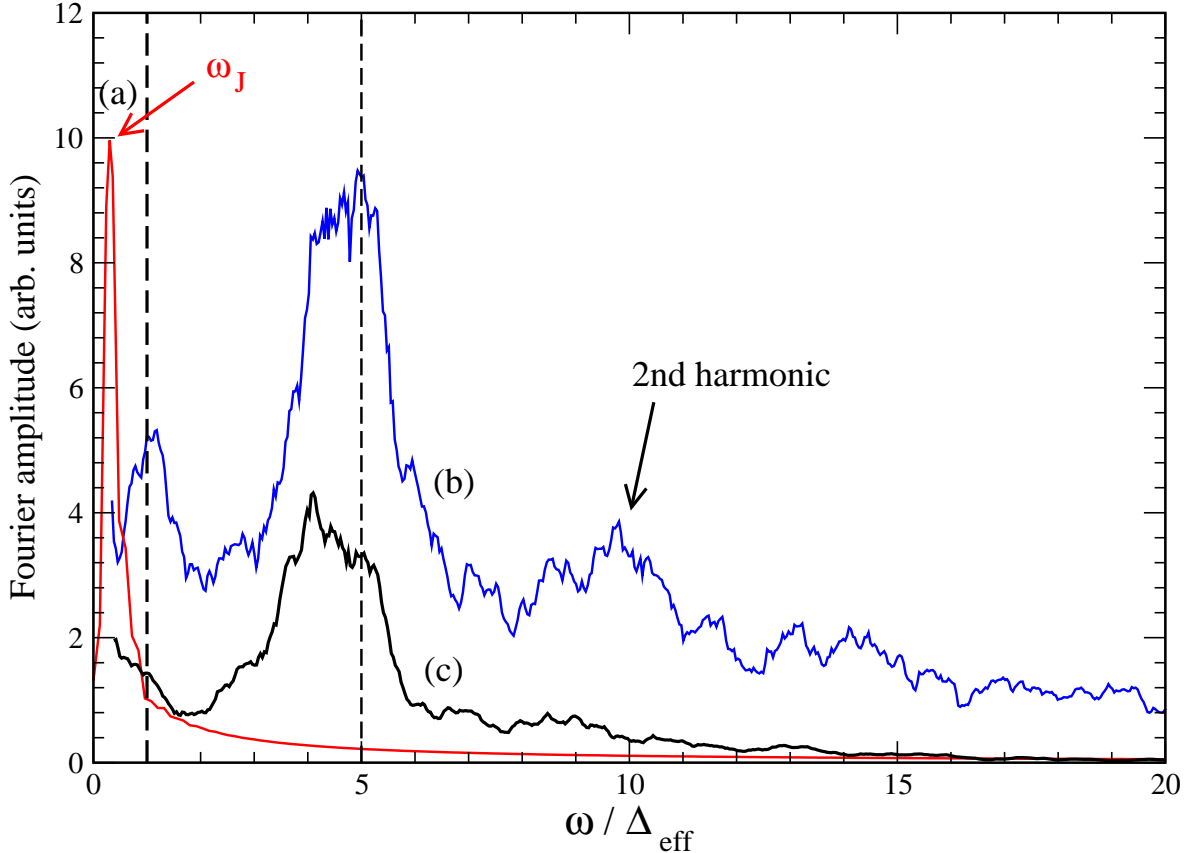}\vspace*{-0.5em}
\end{center}
\caption{
Fourier transforms of the time traces of Fig.~\ref{levels} (b) for (a) red curve: first QP level $n=1$ for $t<\tau_c$,
(b) blue curve: first QP level $n=1$ for $t>\tau_c$ and (c) black curve: condensate mode $\alpha =1$ for $t>\tau_c$. 
The renormalized level spacing $\Delta_{eff}$  and the spacing between the condensate and the uppermost QP level, $5\Delta_{eff}$, 
are marked by vertical dashed lines. 
}
\label{Fourier}
\end{figure}

\begin{figure}[h]
\begin{center}
\includegraphics[width=0.9\textwidth]{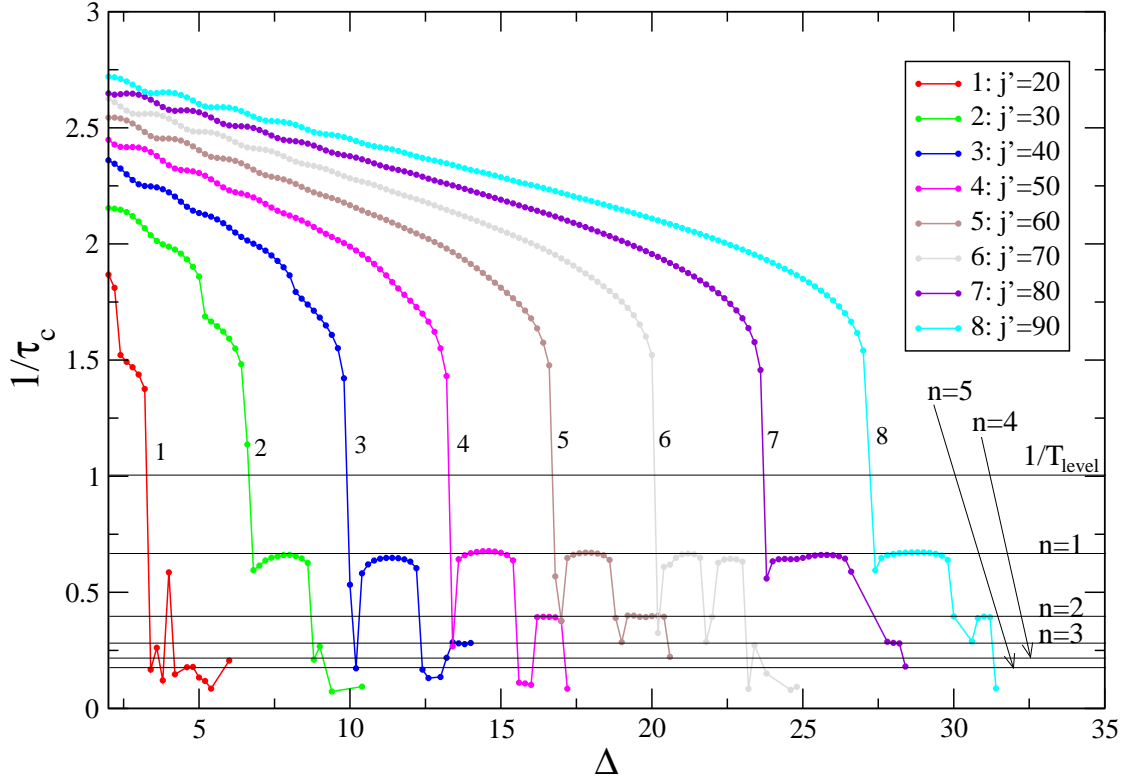}\vspace*{-0.5em}
\end{center}
\caption{
 Dependence of the inverse QP excitation time $\tau_c^{-1}$ on the interlevel spacing $\Delta$ for an initially delocalized Bose Josephson junction ($n(0)=0.6$, $\theta(0)=0$, total number of particles: $N=500000$, $u=u'=5$). Different curves are for different values of the QP-assisted Josephson coupling $j'$. The small dots mark the numerically calculated values. 
The thin horizontal lines depict the discrete values of the 
inverse level crossing times, $t_n^{-1}=2/(2n+1)\, T_{level}^{-1}$ for 
$n=1$, 2, 3, 4, 5 (see text).  
}
\label{alltau_jos}
\end{figure}

\begin{figure}[h]
\begin{center}
\includegraphics[width=0.8\textwidth]{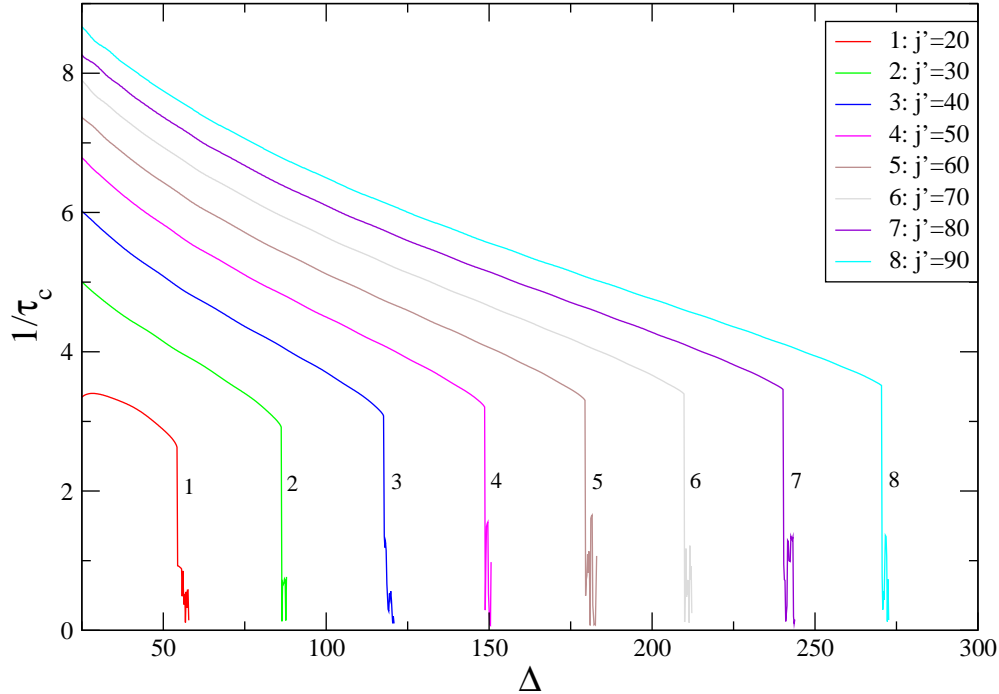}\vspace*{-0.5em}
\end{center}
\caption{
 Dependence of the inverse delocalisation time $\tau_c^{-1}$ on the interlevel spacing $\Delta$ for an initially self-trapped Bose Josephson junction ($n(0)=0.6,  \theta(0)=0$, total number of particles is $N=500000$, $u=u'=25$). Different curves are for different values of the QP-assisted Josephson coupling $j'$. The small dots mark the numerically calculated values.
}
\label{alltau_ST}
\end{figure}

\begin{figure}[!bt]
\begin{center}
\includegraphics[width=0.9\textwidth]{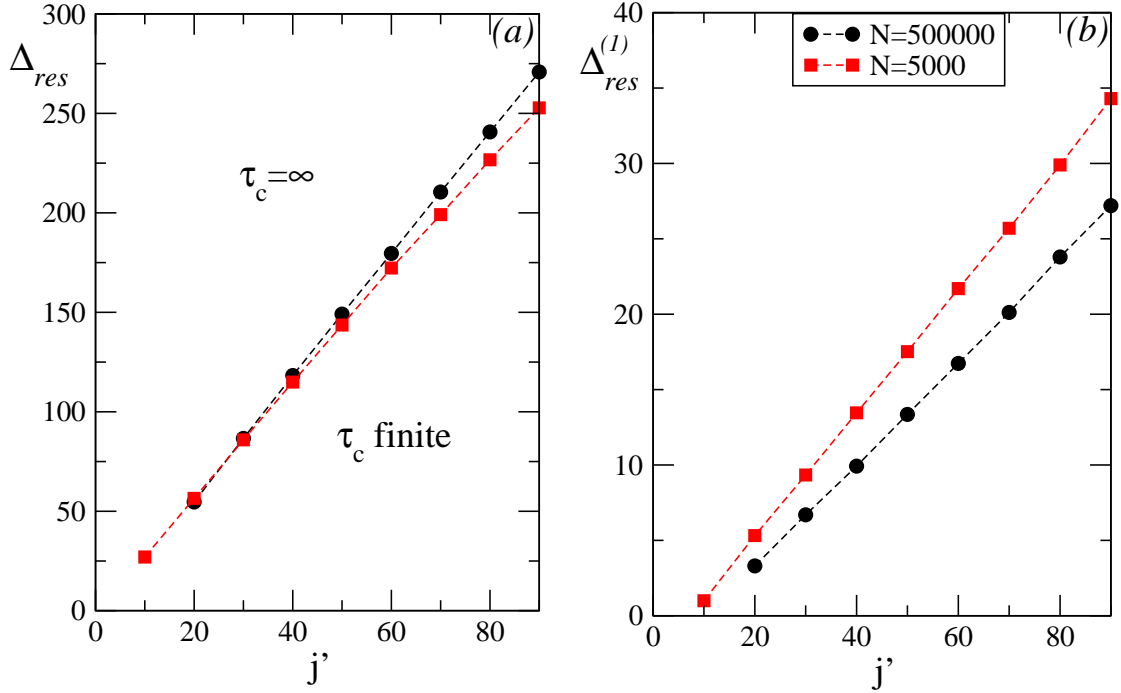}\vspace*{-0.5em}
\end{center}
\caption{
The interlevel spacing $\Delta_{res}$ at which $\tau_c$ first takes a 
discrete value is plotted versus $j'$ for ({a}) an initially self-trapped 
junction ($n(0)=0.6$,  $\theta(0)=0$,  $u=u'=25$) and ({b}) an initially 
delocalized junction ($n(0)=0.6,  \theta(0)=0$,  $u=u'=5$),
each for two values of the particle number $N$.  
}
\label{delta_res}
\end{figure}

\begin{figure}[!bt]
\begin{center}
\includegraphics[width=0.8\textwidth]{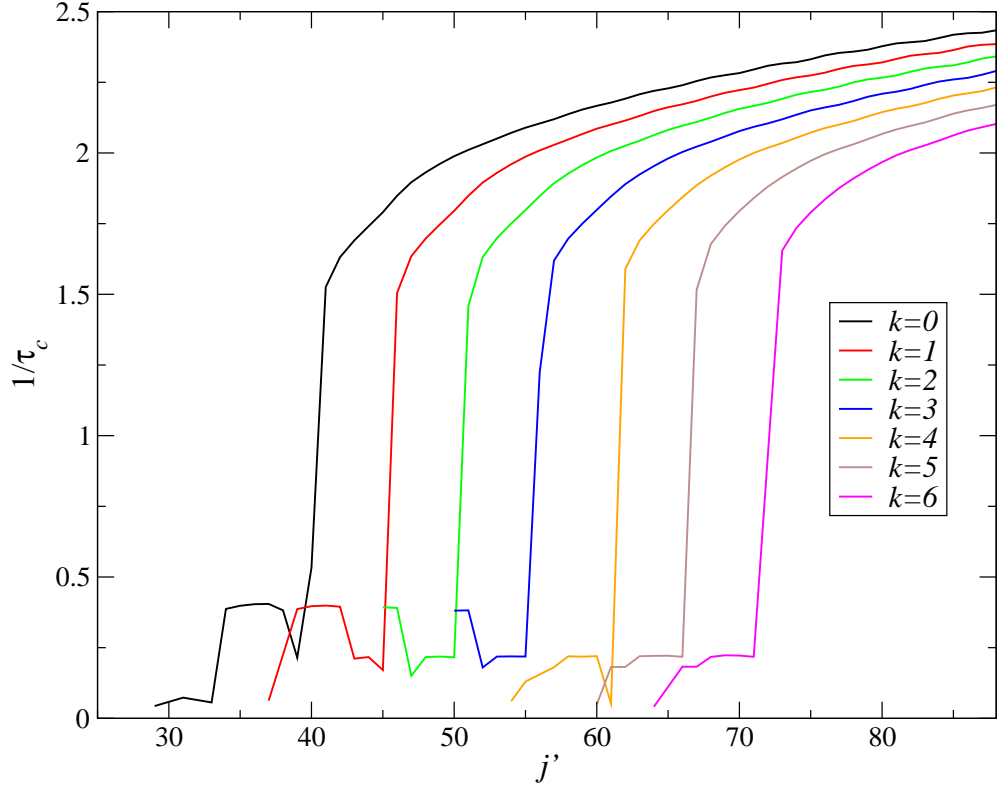}\vspace*{-0.5em}
\end{center}
\caption{
Dependence of $\tau_c^{-1}$ on $j'$ for different values of $k$ and a fixed $\Delta$ ($\Delta=10$). Initially the junction is delocalized: $n(0)=0.6, \theta(0)=0$, $u=u'=5$; total number of particles: $N=500000$. 
}
\label{tau_vs_j}
\end{figure}

\begin{figure}[!bt]
\begin{center}
\includegraphics[width=0.8\textwidth]{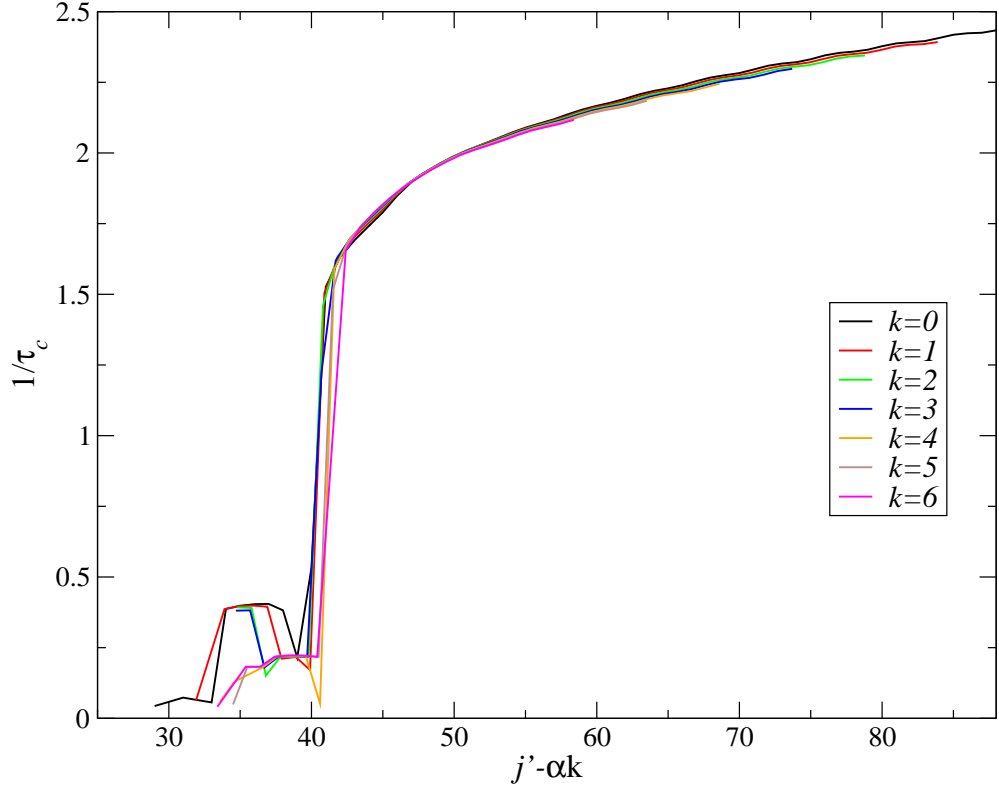}\vspace*{-0.5em}
\end{center}
\caption{
Collapse of the $\tau_c^{-1}(j')$ curves of Fig.~\ref{tau_vs_j}  
by shifting $j' \rightarrow j' -\alpha k$ for different $k$. For the present 
parameters the coefficient $\alpha\approx 5.1$. 
}
\label{tau_vs_jp_scale}
\end{figure}


\section{Results}
\label{Results}

We now solve the system of Eqs. \eqref{self_energy_1} -\eqref{hartree-fock-eq-cond} numerically. We take the initial conditions
\bea
&a_\alpha (0) &= \sqrt{N_\alpha (0)}e^{i\theta_\alpha (0)}, \\
&\text{F}^G _{nm} (0,0) &= -\frac{i}{2}\delta_{nm} ,\\
&\text{F}^F _{nm} (0,0) &= 0, 
\ea
with $z(0)=0.6$ and initial phase difference $\theta_1(0)-\theta_2(0)=0$. The total particle number is taken equal to $N=N_{1}(0)+N_{2}(0)=500000$, with initially all particles in the condensate. It is convenient to define the dimensionless parameters \cite{Mauro09} $u=NU/J$, $k=NK/J$, $j'=NJ'/J$, $u'=NU'/J$, and $n_b ^{(n)}(t) =N_b ^{(n)}(t)/N$ as the relative occupation number of QP level $n$.
For an initially delocalized junction we assume $u=u'=5$ and for an initially self-trapped junction we take $u=u'=25$. We then analyze the behaviour of our system depending on $k$, $j'$ and $\Delta$. In the following, all energies are
expressed in units of $J$, and the time is given in units of $1/J$.

In the absence of coupling to the single-particle excitations ($J'=K=0$) or as long as the QP states are not occupied, the 
equations of motion (\ref{EOMcond}), (\ref{evolution-spectral1}), (\ref{evolution-statistical1}) reduce to 
Eq. (\ref{hartree-fock-eq-cond}) and, hence, for a BEC in a double-well potential, to the well-known 
two-mode approximation for the BEC dynamics \cite{Smerzi97}, exhibiting undamped Josephson oscillations \cite{Mauro09}.
Our main result is, however, the appearance of a new characteristic time scale $\tau_c$, which marks the onset of non-equilibrium QP  dynamics. It may be surprising at first sight that for a discrete level spectrum $\tau_c$ can take a non-zero value \cite{Mauro09}, i.e., QPs are not excited immediately, 
even though the initial state with population imbalance $z(0) >0$ is a highly excited state of the system whose energy exceeds by far the excitation energy of a QP. This is because the BEC Josephson oscillations with frequency $\omega_J \approx 2J \sqrt{1+u/2} $ \cite{Smerzi97} act as a periodic perturbation on the QP system, but QPs cannot be excited in low-order time-dependent perturbation theory, if the Josephson frequency is less than the (interaction-renormalized) level spacing, $\omega_J < \Delta_{eff}$ \cite{Mauro09}. $\tau_c$ depends on system parameters, in particular, the energy splitting $\Delta$ and the QP-assisted Josephson coupling $j'$. We now analyze in detail this new physics.

In Fig.~\ref{imb_jos}({\it a})  the particle imbalance of the initially delocalized junction is shown.  One can clearly distinguish two regimes of slow and fast oscillations, respectively. Comparing with Fig. \ref{imb_jos}({\it c}) reveals that the onset of fast oscillations is marked by an abrupt, avalanche-like population of the single-particle excited states, i.e., it coincides with $\tau_c$. For times $t<\tau_c$ there is only a small, virtual QP population. In this regime, our formalism reduces to the two-mode approximation \cite{Smerzi97} for the condensates, and the dynamics resembles Josephson oscillations. One may, therefore, conjecture that the fast oscillation regime for $t>\tau_c$ is dominated by Rabi oscillations of the single-particle occupation numbers. This will be confirmed below by identifying the oscillation frequencies as the Rabi frequencies for $t>\tau_c$. 
The drastic change is also seen in the phase portrait of the junction in Fig. \ref{imb_jos}({\it b}), where the point on the phase trajectory corresponding to $\tau_c$ is marked by a thick dot. Before $\tau_c$ the time average of the phase difference is equal to zero, as expected in the semiclassical case. However, after $\tau_c$ the time average of the phase difference is equal to $\pi$. We refer to this kind of behaviour as to "$0-\pi$" transition. It may be understood heuristically as follows. One can think of the two-mode BEC system as a driven oscillator, where the QP system provides the driving force. For $t<\tau_c$ this "oscillator" has no driving force and oscillates at its resonance frequency, i.e., the Josephson frequency $\omega_J$ \cite{Smerzi97,Mauro09}, with a phase shift $\theta=0$. However, once in the regime for $t>\tau_c$, the oscillator is periodically driven at the Rabi frequency, far above its resonance frequency, resulting in a phase shift of $\theta=\pi$.

We observe similar behaviour for the initially self-trapped Bose Josephson junction, shown in Fig. \ref{imb_ST}. The scale $\tau_c$ is
substantially smaller in this case (smaller than the period of self-trapped semiclassical oscillations of particle imbalance $T_{ST}\approx 0.6$). At $\tau_c$ the junction becomes delocalized and switches to a fast oscillatory regime with the the time-averaged phase difference equal to $\pi$. The QP occupation numbers show behaviour analogous to that of Fig. \ref{imb_jos}.

To understand the physical origin of the abrupt onset of the QP-dominated regime and the underlying physics of the fast oscillations for $t\ge \tau_c$, we calculate the instantaneous eigenenergies of the system by diagonalizing the Hamiltonian, including the Bogoliubov-Hartree-Fock renormalizations. These eigenenergies are displayed in Fig.~\ref{levels} for an initially delocalized junction. 
Figs. \ref{levels}({\it a}) and ({\it b}) differ only in the value of the bare interlevel spacing $\Delta$. It is seen 
[inset of Fig.~\ref{levels} (b)] that $\tau_c$ occurs at the moment in time when the highest condensate level $E_{c2}$ (shown in red) 
crosses the first (renormalized) QP level $E_1$ for the first time. At this time, QP excitations are no longer blocked energetically, and further QPs are excited resonantly by the fast Rabi oscillations, resulting in an avalanche-like growth of the QP population.  
We emphasize that this is a highly non-perturbative effect. 
The time scale $\tau_c$ increases as the interlevel spacing grows, compare Figs.~\ref{levels} (a) and (b). Eventually, it will become impossible to excite QPs, and $\tau_c$ will tend to infinity. We note in passing that the uppermost (5th) QP level is pushed upward by a level repulsion effect for $t<\tau_c$, but is pulled down by interactions once it gets populated.

The afore-mentioned fast oscillations can now be identified as Rabi oscillations by comparing their frequency with the 
renormalized level spacing $\Delta_{eff}$. To that end we plot in Fig.~\ref{Fourier} the Fourier transforms of the time traces of 
the first QP level $E_1(t)$ for $t<\tau_c$ and $t>\tau_c$, respectively, and also the Fourier transform of the BEC level
$E_{c1}(t)$ for $t>\tau_c$. Similar results are obtained for the other levels.
It is seen that in the QP-dominated regime, $t>\tau_c$, both the condensate and the QP levels oscillate predominantly with frequencies
given by the renormalized level spacing, $\Delta_{eff}$, and with the energy spacing between condensate and the 
uppermost level, $5\Delta_{eff}$. Note that the uppermost level has a large oscillation amplitude, because its oscillation amplitude
is not bounded from above by other levels. This unambiguously shows that the fast oscillations are Rabi oscillations.

We now study the detailed dependence of the characteristic time scale $\tau_c$ 
on 
the level spacing $\Delta$, the QP-assisted tunneling amplitude $j'$ and 
the condensate-QP scattering amplitude $k$.
The dependence of the inverse $\tau_c$ on the interelevel spacing $\Delta$ and QP-assisted Josephson coupling $j'$ for the initially delocalized junction  is shown in Fig. \ref{alltau_jos}. 
First, we note that, before QPs get excited, the oscillation 
period $T_{level}$ of the condensate levels $E_{c1,c2}$, and of the QP 
levels, $E_n$, is half the Josephson oscillation period, 
$T_{level}= 1/2T_J=\pi/\omega_J$, as can be seen from Figs.~\ref{imb_jos} 
and \ref{levels}. This is because the $E_{c1,c2}$, $E_n$ depend on the 
absolute value $|z(t)|$. In Fig.~\ref{alltau_jos} one can distinguish two
regimes of qualitatively different behavior of $1/\tau_c$: For 
$1/\tau_c>1/T_{level}$, $\tau_c$ depends on $\Delta$ in a continuous way, while 
for $1/\tau_c<1/T_{level}$ it jumps between certain discrete or plateau 
values. This is explained as follows. If $\Delta$ is small enough so that 
$1/\tau_c>1/T_{level}$, the condensates cannot perform a full Josephson 
oscillation before QPs get excited. Therefore, the BEC dynamics cannot 
be considered a periodic perturbation on the QP system, and any value of 
$\tau_c$ is possible, thus increasing continuously with $\Delta$. 
For $1/\tau_c<1/T_{level}$, however, the QP system gets excited when one of the
BEC levels $E_{c1,c2}$ and the first QP level $E_1$ come close to 
each other and eventually cross, see Fig.~\ref{levels}. In the 
time-periodic regime this occurs for the discrete times 
$t_n=(n+1/2)T_{level}$, $n=1,\ 2,\ \dots$. Therefore, the $t_n$ mark 
preferred values for the QP excitation time $\tau_c$. 
In Fig.~\ref{alltau_jos} the $1/t_n$ are marked as thin, horizontal lines  
$1/\tau_c$ jumps discontinuously between those values, with plateau-like   
behavior between the jumps. Which of the $1/t_n$ is realized for
a particular level spacing $\Delta$ depends on the detailed non-linear
dynamics of the system and can lead to resonance-like behavior, as seen in
Fig.~\ref{alltau_jos}. For $n\to \infty$, $\tau_c$ diverges. Since the 
numerical time evolution is limited to finite times, we can resolve only 
finite $n$.    
For the initially self-trapped junction, one observes similar behavior
(Fig.~\ref{alltau_ST}), however 
for substantially greater values of $\Delta$, due to the substantially 
larger BEC oscillation frequency (compare Figs.~\ref{imb_jos} 
and \ref{imb_ST}).

The smallest value of the level spacing, $\Delta_{res}$, for which a 
discrete $\tau_c=t_n$ is realized (resonance-like structures in 
Fig.~\ref{alltau_jos}) can be used to separate the regimes of continuous 
and discrete $\tau_c$ behavior. We find a linear increase of $\Delta_{res}$ 
with the QP-assisted Josephson tunneling amplitude $j'$, as seen in 
Fig.~\ref{delta_res}. In the regime below the line $\tau_c$ behaves 
continuously, above the line it takes discrete values.   
Note, that this ``phase diagram'' depends only weakly on the number of 
particles in the system.

In Fig. \ref{tau_vs_j} we present the dependence of the inverse $\tau_c$ 
versus QP-assisted coupling $j'$ for various values of the
QP-BEC interaction $k$ for an initially delocalized junction. 
One can see, when $k$ is non-zero, it becomes more difficult to excite 
QPs, and increasing $k$ results in larger values of $\tau_c$.  
We attribute this remarkable behavior to the fact that $k$ implies a 
level repulsion between the BEC levels $E_{c1,c2}$ and the QP level $E_1$ 
and, hence, inhibits QP excitations. At the resonance point, where  
$E_{c2}$ and $E_1$ cross, the k-induced level repulsion is linear in $k$.
Therefore, one expects $\tau_c$ to increase linearly with $k$. 
Indeed, the curves $\tau_c^{-1}(j')$ for different $k$ collapse onto a 
single curve by shifting $j'$ by a term linear in $k$, as shown in 
Fig. \ref{tau_vs_jp_scale}.

\section{Conclusions and Discussion}

\label{Conclusions}

We studied in detail the non-equilibrium dynamics of a double-well Bose Josephson junction in two cases: the junction is initially delocalized, and the junction is initially self-trapped in a semiclassical meaning described in Ref. \cite{Smerzi97}. The non-equilibrium is coursed by the abrupt switching on of the Josephson coupling between the well, which corresponds to the experimental situation \cite{Albiez05}. We treat our system within Bogoliubov-Hartree-Fock first order approach and develop within it the non-equilibrium equations of motion for the condensate and QP parts of our Hamiltonian.

Our main finding is that the quasiparticle dynamics does not immediately set in at the time when the Josephson coupling is turned on, but after a certain period $\tau_c$. At $\tau_c$, QPs are excited in an avalanche manner due to fast quasiparticle oscillations between the discrete energy levels of the system. $\tau_c$ corresponds to a moment in time when the highest condensate eigenenergy is coincident with the eigenenergy of the first excited level. It depends on the system parameters in a complex way, including regimes of continuous 
and discrete behavior, which we have analyzed in the present work. 
The appearance of a regime with large, discrete $\tau_c$ explains the 
experimental finding that a junction can maintain undamped Josephson 
oscillations, although the barrier between the walls was ramped up in a 
sudden way \cite{Albiez05}.

We find furthermore that, when QP dynamics sets in, the junction inevitably switches to a $\pi$ junction, a junction, whose time-dependent phase difference is equal to $\pi$ when averaged over time. This is a necessary condition for sustainig fast Rabi-like oscillations between the QP levels. In the future we aim at studying how the non-equilibrium Josephson junction will equilibrate (if at all). This requires considering collisions between the quasiparticles in the full second order approximation.


\newpage

\section*{Appendix: Particle and energy conservation}

\label{conservation}
For any closed system the particle number and energy are conserved quantities. The particle number conservation arises as a consequence  of the invariance of the Hamiltonian under global phase change. The mean particle number can be expressed in terms of the condensates wavefunctions and the statistical function, as follows
\bea
\la N \ra (t)= N_1 + N_2 + \sum_{n\ne 0} N_b ^{(n)} 
=\sum_{\alpha =1,2} a^* _\alpha (t) a_\alpha (t) + \sum_{n\ne 0} \left[i\text{F}^G _{nn} (t,t) - \frac{1}{2} \right]
\ea
and it can be proven to be constant by making use of the equations of motion for $a_\alpha$ and $F^G _{nn}$. 
It is well known, that the exact solution of an isolated system excludes dissipation. One must therefore take care that the self-energies are derived within a conserving approximation (otherwise the non-physical dissipation can take place). If (like in our case) the self-energies ${\bf S}$ and ${\bf \Sigma}$ are derived by means of a conserving approximation, for instance from a Luttinger-Ward functional \cite{Baym61,Baym62}, the mean energy can be written as 
\beq
\la H \ra = \la H \ra_{cond} + \la H \ra_{exc},
\eq
and can be proven to be conserved. The energy of the condensate fraction is 
\bea
\la H \ra_{cond}(t) = \frac{i}{2}  \mathrm{Tr}\left[\left(E_{\alpha\beta}\mathds{1} +\frac{1}{2}{\bm S}^{HF}_{\alpha\beta}(t)\right){\bm C}_{\beta\alpha}(t,t) \right]
+\frac{1}{2}  \mathrm{Tr}\left[ \int\limits_{-\infty}^{t} d \ov{t}  {\bm \gamma}_{\alpha\gamma}(t,\ov{t}){\bf C}_{\gamma\beta} (\ov{t} , t)\right],
\ea
and the energy of the QP excitations is
\bea
\la H \ra_{exc}(t) = \frac{i}{2}  \mathrm{Tr}\left[\left(\epsilon_{n}\delta_{nm}\mathds{1} +\frac{1}{2}{\bm \Sigma}^{HF}_{nm}(t)\right){\bf F}_{mn}(t,t) \right]\nn \\
+\frac{1}{2}  \mathrm{Tr}\Big{[}  \int\limits_{-\infty}^{t} d \ov{t}  \big{(}{\bm\Gamma}_{nm}(t,\ov{t}){\bf F}_{mn}(\ov{t},t)
- {\bm \Pi}_{nm}(t,\ov{t}){\bm A}_{mn}(\ov{t},t)\big{)} \Big{]}\nn\\
+\frac{1}{2}  \mathrm{Tr}\Big{[}  \int\limits_{-\infty}^{t} d \ov{t}  \big{(}{\bf F}_{nm}(t,\ov{t}){\bm \Gamma}_{mn}(\ov{t},t) 
- {\bm A}_{nm}(t,\ov{t}){\bm \Pi}_{mn}(\ov{t},t)\big{)} \Big{]}.
\ea
Here again we sum over all indices appearing twice in every term.
In the numerical solution of the time-dependent equations of motion
the self-energies are calculated inherently in a self-consistent way,
since in each time step of the numerical solution, the full solution of the 
previous step is used.



\newpage



\begin{thebibliography} {99}
\bibitem{footnote}
This is exactly true if the left and right wells are separated.
\end{thebibliography}
\end{document}